# Direct evidence for flat bands in twisted bilayer graphene from nano-ARPES

*Simone Lisi[1†], Xiaobo Lu[2†], Tjerk Benschop[3†], Tobias A. de Jong[3†], Petr Stepanov[2], Jose R. Duran[2], Florian Margot[1], Irène Cucchi[1], Edoardo Cappelli[1], Andrew Hunter[1], Anna Tamai[1], Viktor Kandyba[4], Alessio Giampietri[4], Alexei Barinov[4], Johannes Jobst[3], Vincent Stalman[3], Maarten Leeuwenhoek[3,5], Kenji Watanabe[6], Takashi Taniguchi[6], Louk Rademaker[7], Sense Jan van der Molen[3], Milan Allan[3], Dmitri K. Efetov[2], Felix Baumberger[1,8]*

[1]*Department of Quantum Matter Physics, University of Geneva, 24 Quai Ernest-Ansermet, 1211 Geneva 4, Switzerland*
[2]*ICFO – Institut de Ciencies Fotoniques, The Barcelona Institute of Science and Technology, Castelldefels, Barcelona, Spain*
[3]*Huygens-Kamerlingh Onnes Laboratory, Leiden Institute of Physics, Leiden University, Niels Bohrweg 2, 2333 CA Leiden, The Netherlands*
[4]*Elettra-Sincrotrone Trieste S.C.p.A., Basovizza, 34149 Trieste, Italy*
[5]*Kavli Institute of Nanoscience, Delft University of Technology, Lorentzweg 1, 2628CJ Delft, The Netherlands.*
[6]*National Institute for Materials Science, 1-1 Namiki, Tsukuba, 305-0044, Japan*
[7]*Department of Theoretical Physics, University of Geneva, 24 Quai Ernest-Ansermet, 1211 Geneva 4, Switzerland*
[8]*Swiss Light Source, Paul Scherrer Institute, CH-5232 Villigen PSI, Switzerland*

[†]*These authors contributed equally to this work*

**Transport experiments in twisted bilayer graphene revealed multiple superconducting domes separated by correlated insulating states[1–5]. These properties are generally associated with strongly correlated states in a flat mini-band of the hexagonal moiré superlattice as it was predicted by band structure calculations[6]. Evidence for such a flat band comes from local tunneling spectroscopy[7–11] and electronic compressibility measurements[12], reporting two or more sharp peaks in the density of states that may be associated with closely spaced van Hove singularities. Direct momentum resolved measurements proved difficult though[13]. Here, we combine different imaging techniques and angle resolved photoemission with simultaneous real and momentum space resolution (nano-ARPES) to directly map the band dispersion in twisted bilayer graphene devices near charge neutrality. Our experiments reveal large areas with homogeneous twist angle that support a flat band with spectral weight that is highly localized in momentum space. The flat band is separated from the dispersive Dirac bands which show multiple moiré hybridization gaps. These data establish the salient features of the twisted bilayer graphene band structure.**

The small rotational misalignment of the sheets in twisted bilayer graphene (TBG) results in a long-range moiré superstructure with a unit cell containing several thousand atoms. Moiré mini-bands leading to physical properties that deviate strongly from those of aligned bilayer graphene can form in structurally highly perfect devices where electronic states are coherent over multiple moiré unit cells and thus over a great number of atomic sites. The formation of mini-bands further requires a finite overlap of the low-energy orbitals between neighboring moiré sites, which in turn suggests extended wave functions with a weak onsite Coulomb repulsion. Nevertheless, near the magic twist angle of ≈1.1°, TBG shows hallmarks of electron-electron correlations such as metal-insulator transitions, magnetism[3,4], superconductivity[2,3,5] and departures from Fermi liquid behavior in the metallic state[14] that are more commonly observed in 3$d$ transition metal oxides with on-site interactions of several eV.

This can be reconciled by assuming a marked flattening of the dispersion in the moiré mini-bands, as predicted by band structure calculations. Describing the electronic structure and related many-body physics of TBG is theoretically demanding given the size of its unit cell. It is thus important to test key-predictions of band structure calculations experimentally. This proved challenging though and direct electronic structure measurements by ARPES have thus far largely been limited to macroscopic samples of epitaxially grown bilayers with large and uncontrolled twist angles. Such measurements have shown signatures of the superlattice periodicity[15] and flat bands deep in the occupied states[16] but did not show evidence for the predicted partially filled flat band that is believed to be responsible for the correlated behavior of TBG near the magic angle. Evidence for the latter has been reported in a very recent room temperature study on a device made from exfoliated graphene that was strongly influenced by an additional moiré superlattice arising from a small twist angle with the hexagonal boron nitride (hBN) substrate[13].

Here we provide direct evidence for the existence of flat bands in TBG near the magic angle. This is achieved by combining low-energy electron microscopy (LEEM) and scanning tunneling microscopy (STM) with nano-ARPES, a technique that can image the photocurrent with sub-micron spatial resolution while providing simultaneous and fully independent momentum space resolution.

Multiple TBG devices were fabricated by the tear and stack method, as described in Supplementary Information. As shown in the schematic in Fig. 1a, the two graphene monolayers are supported by a hexagonal boron nitride flake isolating the structure from a graphite electrode. The latter is connected to a prepatterned Au contact on a Si/SiO$_2$ substrate. A graphite stripe is used to connect the TBG to the second electrode. The TBG and graphite bottom electrode were both grounded for all experiments.

The key difference to TBG devices used for transport experiments is the absence of a hBN flake encapsulating the structure. This allows unimpeded access for surface techniques but poses a challenge for device fabrication. In particular, it has thus far not been possible to make such open TBG devices with a twist angle homogeneity that rivals encapsulated devices[3,17]. Neither has the actual twist angle of such devices been determined from gate dependent transport experiments. This is a serious obstacle for nano-ARPES experiments since the twist angle of devices frequently changes during fabrication and therefore cannot be reliably predicted. Finally, the electronic properties of open devices are more susceptible to the degrading effect of polymer residues and hydrocarbon contamination of the surface. A thorough characterization of the twist angle and cleanliness of the devices prior to ARPES experiments is thus essential.

This is achieved here by combining low-energy electron microscopy (LEEM) and scanning tunneling microscopy (STM). In Fig. 2b, we use bright field LEEM for a large-scale characterization of the area in which the two twisted graphene monolayers overlap. This shows a large area free of folds and bubbles of gases trapped between the layers, but with several round features with lateral dimensions of typically 2 μm, which we associate with agglomerates of polymer residues.

Contrast differences between areas are attributed to different local lattice stackings[18]. By combining with the dark field LEEM overview in Fig. 2c, we use this stacking contrast to classify areas. As dark field imaging shows strong contrast between AB and BA stacked Bernal graphene, we can unambiguously identify the large, homogeneous, intermediate intensity area in Fig. 2c as TBG, corresponding to slightly lower intensity in Fig. 2b. It is separated by small folds, visible as dark straight lines, from areas that have reconstructed into Bernal stacking. Some of these areas exhibit alternating AB and BA stacked triangles and can thus be identified as TBG with very small twist angle[19]. The smallest such structures that we can resolve have a line pitch of 25 nm, corresponding to a twist angle of 0.55°. As no such structures can be observed in the homogeneous TBG area, this also gives us a lower bound on the twist angle there of 0.55°. Furthermore, we obtain an upper bound of 2° from the absence of satellite peaks of the moiré periodicity in micro low-energy electron diffraction (μ-LEED) and the momentum space resolution of these experiments. The same μ-LEED diffraction patterns confirm an angle of 29±1° between TBG and hBN. We have deliberately chosen such a large angle to reveal the intrinsic electronic structure of TBG by minimizing competing moiré effects from the interaction with hBN.

For a more precise determination of the twist angle in bilayer graphene, we use STM topographic images acquired on the same device (Fig. 2d). These images show moiré periodicities of 10.3 to 10.8 nm corresponding to a variation of the twist angles between 1.31° and 1.37° over a distance of approximately 1μm probed by these experiments. Further evidence for a good homogeneity of the TBG areas comes from position dependent nano-ARPES experiments. Two representative dispersion plots acquired at different positions on the same TBG area are shown in Fig. 2e and show excellent reproducibility. All main features do not change noticeably between these two spots. This is a prerequisite for the reliable acquisition of the detailed 3D ARPES data sets, which we discuss in the following.

The application of the concept of band structures is not quite straightforward for TBG. Only for a discrete set of twist angles TBG is translationally invariant and thus a crystal that supports Bloch states in a strict sense[6]. In the general incommensurate structure, the spectrum of eigenvalues is dense at every momentum, fundamentally different from the continuous $E(k)$ dispersion relation typical of electrons in simple crystals. Yet, experiments on TBG show clear evidence for band-like transport at any twist angle, irrespective of whether the structure is commensurate or not[2,3,14]. This can be understood by supposing the formation of a quasi-band-structure from the non-uniform distribution of spectral weights over the complex eigenvalue spectrum, as was proposed for incommensurate density wave systems[20]. ARPES directly measures these spectral weights[20,21]. In Fig. 3a we indeed find that the photoemission intensity at the Fermi level is highly localized near the $K_1$ and $K_2$ points of the two twisted monolayers from where it disperses away with increasing energy. The spectral weight thus singles out a small subset of all possible low-energy eigenvalues, providing direct support for the emergence of Bloch-like bands out of the complex spectrum of eigenvalues in a moiré structure and thus for the widely used continuum models of the band structure.

The effect of the twist angle on the details of the band structure is profound. In Fig. 3b-d, we show a series of constant energy cuts and compare them to band structure calculations of the spectral weight for an isolated TBG layer with a twist angle of 1.34°. At all energies, we find a far more complex electronic structure than in Bernal bilayer graphene, where constant energy contours are simple concentric circles with small trigonal warping. The electronic structure observed in experiment is also fundamentally different from bilayer graphene with large twist angle where constant energy contours are well described by two weakly hybridized circles centered at $K_1$ and $K_2$, respectively[22]. On TBG instead, we find a complex spectral weight texture with multiple contours that appear to be centered at the $\Gamma$ points of the 4 mini-Brillouin zones surrounding $K_{1,2}$. This is particularly evident in the calculation at -0.2 eV but similar features can be recognized at higher energy and in the data too. We interpret these $\Gamma$-centered constant energy contours as the result of hybridization of Dirac cones at all K-points of the moiré mini-Brillouin zone. This naturally results in band extrema at $\Gamma$ rather than at $K_{1,2}$, as observed for large twist angles. Our observation of such structures thus provides direct evidence for the strong interlayer coupling and the formation of moiré mini-bands predicted for small angle TBG[6,23]. We note that some Fermi surfaces appear broken into apparent arc like structures that seemingly end at arbitrary momenta. Importantly though, these structures likely do not represent genuine Fermi arcs as those observed in Weyl semimetals[24] and possibly in cuprates[25]. A detailed inspection of the calculations rather suggests that they emerge from a multitude of very small but closed surfaces whose spectral weight decays away from $K_{1,2}$ but remains finite (see supplementary Fig. 3).

In Fig. 4 we focus on the *E(k)* dispersion relation of the quasi-Bloch bands of TBG. The overview of cuts perpendicular to the $K_1$-$K_2$ line in panel 4a already reveals the predicted dichotomy of the electronic states with two distinct sub-systems. Most importantly, the raw data directly show a flat band with spectral weight localized near the $K_{1,2}$ points that is separated from the dispersive bands. The latter can, to first approximation, be attributed to the $K_1$ and $K_2$ Dirac cones split by ~4*u'* along this *k*-space direction, where *u'* is the interlayer coupling (see Supplementary Information). The detailed comparison of selected cuts with calculations of the spectral weight distribution in Fig. 4c-e reveals additional richness. First, we find clear evidence for hybridization gaps in the dispersive bands reflecting the moiré superlattice potential, as indicated by the black arrows in panel d. The dominant gaps have magnitudes up to ≈150 meV and are thus comparable to *u'*. We note that the calculations further predict a multitude of smaller gaps from higher order Umklapps. These cannot presently be resolved in our data.

The spectral weight at the Fermi level corresponds to a flat band, as shown in Fig. 4c-e. Despite its localized spectral weight in *k*-space, which reflects the extended nature of the wavefunctions in real space, we can estimate that its group velocity near the K-point of the mini-Brillouin zone is around an order of magnitude lower than the velocity in monolayer graphene. This estimate is in fair agreement with our calculations showing a renormalization of the group velocity in the moiré Dirac cone of 0.19 compared to monolayer graphene. More precise measurements of the mini-bandwidth for different twist angles might become possible in future nano-ARPES experiments with improved resolution following the procedures outlined here. We note that the flat band is clearly separated from the Dirac bands for most of its extension in *k*-space. For certain cuts, however, it appears to touch the latter within the resolution of the experiments. From this we estimate an upper limit of the gap between these 2 subsystems of ≈50 meV, consistent with calculations that incorporate the effect of structural relaxation in the bilayer[26,27].

In conclusion, we report direct measurements of the electronic structure of TBG near the magic twist angle using nano-ARPES on carefully prepared devices. Our data directly confirm the salient features of the TBG band structure that were thus far deduced from transport measurements and theory. This holds in particular for the separation between dispersive high energy bands and a flat band at the Fermi level predicted from continuum models. This provides a basis for further theoretical work and establishes the potential of nano-ARPES for electronic structure studies on devices showing highly non-trivial physics.

**Methods:**

**Fabrication.** A schematic of the fabrication processes is shown in supplementary Fig. 1. First, HBN flakes were exfoliated on a PDMS stamp. Then two graphene flakes on a $SiO_2$/Si substrate were sequentially picked up with hBN on PDMS. The two pieces of graphene came from a single graphene flake that was pre-cut with an AFM tip and were manually twisted 1.3°. To avoid that the hBN flake dropped down on the $SiO_2$/Si substrate instead of picking up graphene, the hBN flake was always kept partially contacted with the substrate during the picking up processes (as shown in supplementary Fig 1b-c). Subsequently, the TBG/hBN structure was flipped over and picked up with a second PDMS stamp and then transferred onto a graphite flake which is pre-transferred on the $SiO_2$/Si substrate and connected with a Au electrode as a gate. Finally, a second piece of graphite was placed between TBG and a second pre-patterned Au electrode as a contact.

Prior to the measurements shown here, samples have been annealed at ~350°C in ultrahigh vacuum for several hours.

**Nano-ARPES** experiments were performed at the SpectroMicroscopy beamline of Elettra light source[28]. This instrument uses multilayer coated Schwartzschild objectives with a numerical aperture of 0.2 to de-magnify a pinhole located at an intermediate focus on the sample and achieves a spatial resolution of ~600 nm. All experiments were performed at T = 85 K with *p*-polarized light with a fixed incidence angle of 45°. K-space mappings were performed by rotating an imaging hemispherical analyzer mounted on a 5-axes goniometer. The combined energy and momentum resolution of the experiments was ~ 45 meV / 0.005 Å$^{-1}$.

**LEEM.** Before PEEM and LEEM imaging, samples were annealed at 350°C, as measured by pyrometer. Imaging was performed at the same temperature to prevent beam contamination. Images were recorded in HDR mode and corrected for detector artefacts as described in Ref. [29]. PEEM imaging was performed using an unfiltered mercury short-arc lamp with its main emission at a photon energy of ~6 eV. Dark Field imaging was performed under tilted illumination, as described in detail in Ref. [30] Furthermore, overviews were stitched together using a cross-correlation based method and globally intensity matched.

**STM measurements.** The devices were inserted in our home-built, low-temperature (4.2 K), UHV (< 3.0 x 10$^{-10}$ mbar) setup, featuring a commercial STM head from *RHK inc.* and a cryostat *CryoVac GmbH*. The devices were then annealed to 350°C for approximately 10 hours before insertion into the STM head. To land the STM tip on the TBG sample, we utilize the capacitive navigation method described in Ref. [31]. Our Si/SiO$_x$ chip contains a patterned gold contact, on which we apply an AC voltage of $V_{pp}$ = 1 V at 5 kHz with respect to ground. The same signal, but rotated 180° out of phase is applied to the Si chip. We then use the coarse motor to move the tip laterally at a few micrometers above the sample, and use the strength of the capacitive signal to guide the movement with respect to the gold pattern. Once the TBG flake is located, we approach the tip to perform the STM measurements. All STM measurements were performed with mechanically polished PtIr tips from Unisoku.

**Extraction of the twist angle.** To determine the twist angle, we Fourier transform topographic images of 113x113nm$^2$, and measure the distance |q$_0$| in q space to each moiré peak (supplementary Fig. 2).

The wavelength $\lambda_M$ of the Moiré lattice is then determined by calculating the moiré wavelength $\lambda_M = \frac{4\pi}{\sqrt{3}|q_0|}$. Finally, the twist angle $\theta$ is obtained through the relation $\lambda_M = \frac{a}{2\sin\frac{\theta}{2}}$, where a = 0.246 nm, is the graphene lattice constant and $\theta$ is the twist angle.

**Calculations.** To compute the theoretical ARPES intensity, we modeled each layer using a nearest-neighbor tight-binding model and the interlayer coupling using the standard twisted continuum theory. The ARPES intensity was obtained by projecting the electron wavefunctions for twist angle θ= 1.34° onto the first mini-Brillouin zone. Details can be found in the supplementary information.

**Acknowledgements**

We thank J. Aarts, S. Nadj-Perge, A. Yazdani, A. Pasupathy, A. Morpurgo, I. Gutierrez-Lezama, H. Henck, F. Groenewoud, K. van Oosten, R. Wijgman and H. Zandvliet for discussions. We thank M. Hesselberth for technical LEEM support. The ARPES work was supported by the Swiss National Science Foundation (SNSF) through grant 200020_184998. L.R. acknowledges support by the SNSF through an Ambizione grant. The STM work was supported by the European Research Council (ERC StG SpinMelt) and by the Dutch Research Council (NWO), as part of the Frontiers of Nanoscience programme, as well as through a Vidi grant (680-47-536). The LEEM work was supported by the Netherlands Organization for Scientific Research (NWO) as part of the Frontiers of Nanoscience program. Growth of hexagonal boron nitride crystals was supported by the MEXT Element Strategy Initiative to Form Core Research Center (JPMXP0112101001) and the CREST (JPMJCR15F3), JST. D.K.E. acknowledges support from the Ministry of Economy and Competitiveness of Spain through the "Severo Ochoa" program for Centres of Excellence in R&D (SE5-0522), Funda-ció Privada Cellex, Fundació Privada Mir-Puig, the Generalitat de Catalunya through the CERCA program, the H2020 Programme under grant agreement n° 820378, Project: 2D·SIPC and the La Caixa Foundation.

**Data availability**

The data relevant to the findings of this study are available from the corresponding authors on reasonable request.

**Author contribution**

XL, PS and JRD made the TBG devices. TT and KW contributed hBN materials. SL, FM, IC, EC and AH performed the nano-ARPES experiments. TB, VS and ML performed STM experiments. TdJ acquired the LEEM and μ-LEED data. LR performed the band structure calculations. JJ, SJvdM (LEEM), MA (STM), DKE (devices) and FB (nano-ARPES) were responsible for the project direction and the provision of resources. VK, AG and AB were responsible for the nano-ARPES beamline. SL, AT and FB wrote the bulk of the manuscript with contributions from several others. All authors contributed to the scientific discussion of the results.

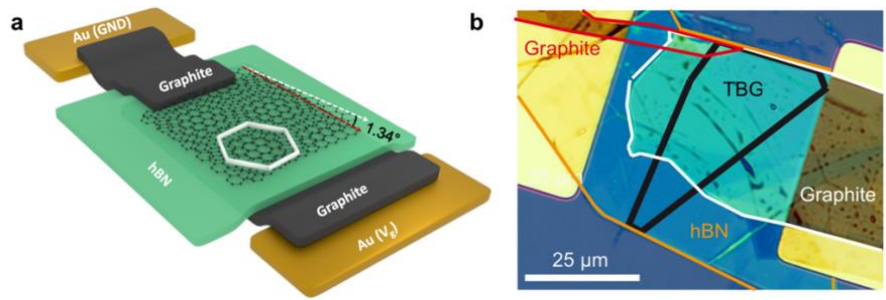

**Figure 1 | Device layout**. **a** Sketch of the van der Waals stack with TBG on top of hBN and a bottom graphite electrode. The TBG is contacted by a graphite stripe. **b** Optical micrograph of the device with boundaries of the different layers indicated as guide to the eye. Details of the device fabrication are given in methods and Supplementary Information.

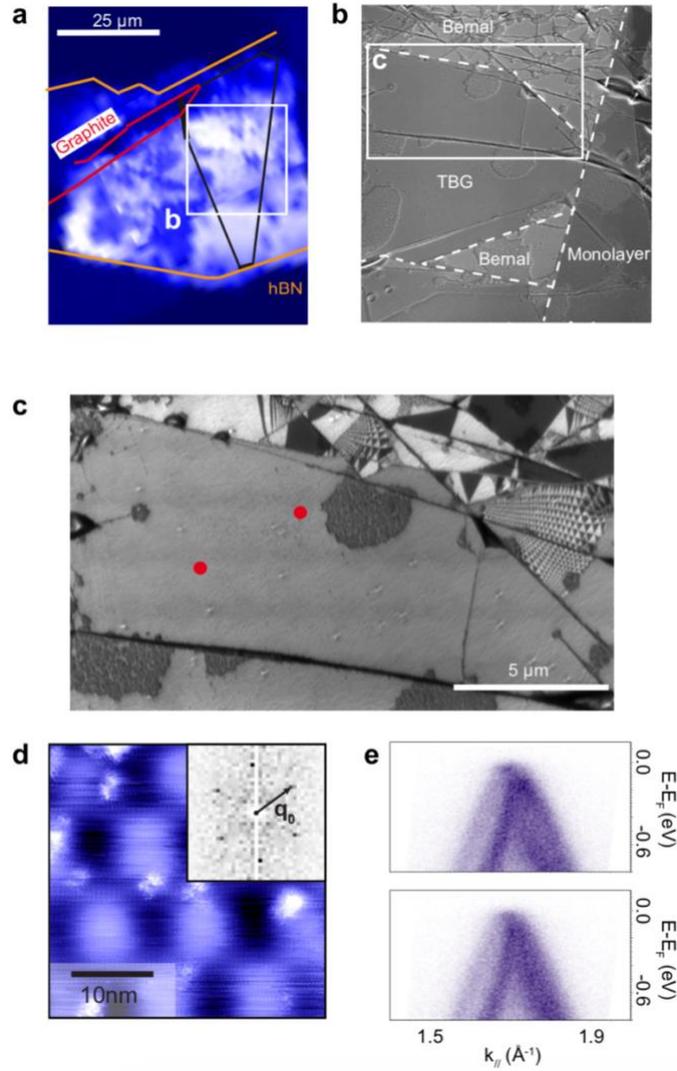

**Figure 2 | Device characterization. a** Intensity map of the ARPES signal integrated over 3.3 eV from the chemical potential along a *k*-space cut crossing the K-point of one of the graphene layers in TBG. **b** Bright field LEEM image at a landing energy of 0.2 eV of the area indicated in panel **a**. Boundaries between different stacking contrasts are indicated as a guide to the eye. **c** Dark field image at a landing energy of 28.0 eV of the area indicated in **b**. Triangular reconstruction of low twist angle is visible on the right. **d** STM topography acquired with set-up voltage and current of V = -250 mV, I = 100 pA. Inset: Fourier transform used for the determination of the twist angle, as described in methods. **e** ARPES measurements at the positions indicated in panel **c** showing good homogeneity of the sample.

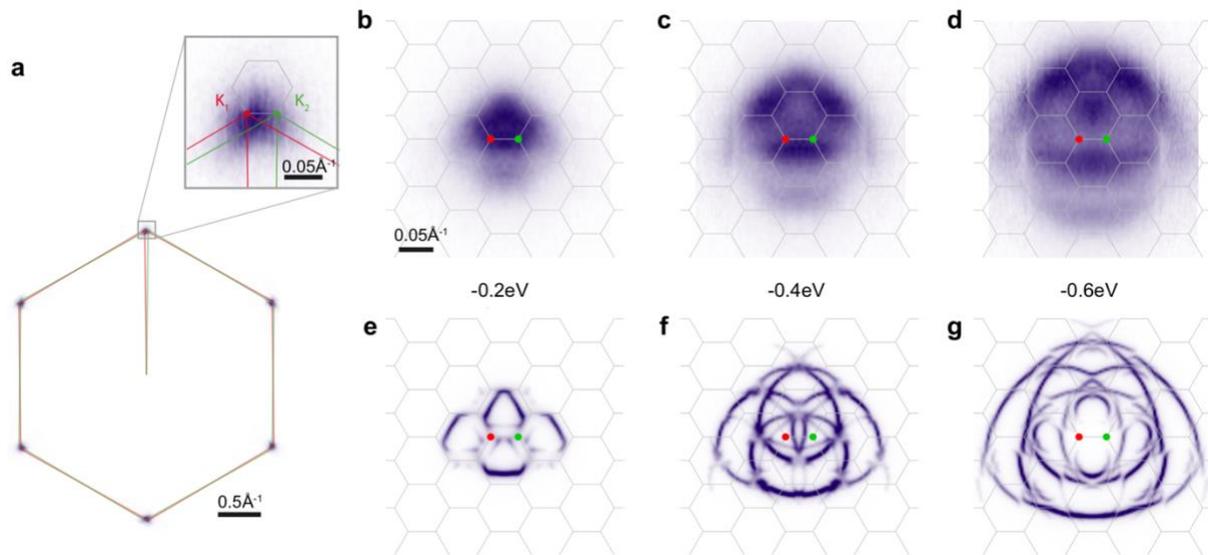

**Figure 3 | ARPES spectral weight distribution. a** Spectral weight at the Fermi level over the full Brillouin zones of the two graphene monolayers (red, green). The inset shows a zoom-in near the $K_{1,2}$ points of the two layers (marked in all panels by red and green points). **b-d** ARPES constant energy contours at -0.2 eV, -0.4 eV and -0.6 eV. The data have been symmetrized along the vertical axis. The hexagonal tiles are moiré mini Brillouin zones. **e-g** Calculations of the spectral weight within the continuum model. A Lorentzian broadening of 10 meV has been applied to the calculation. For details of the calculations, see Supplementary Information.

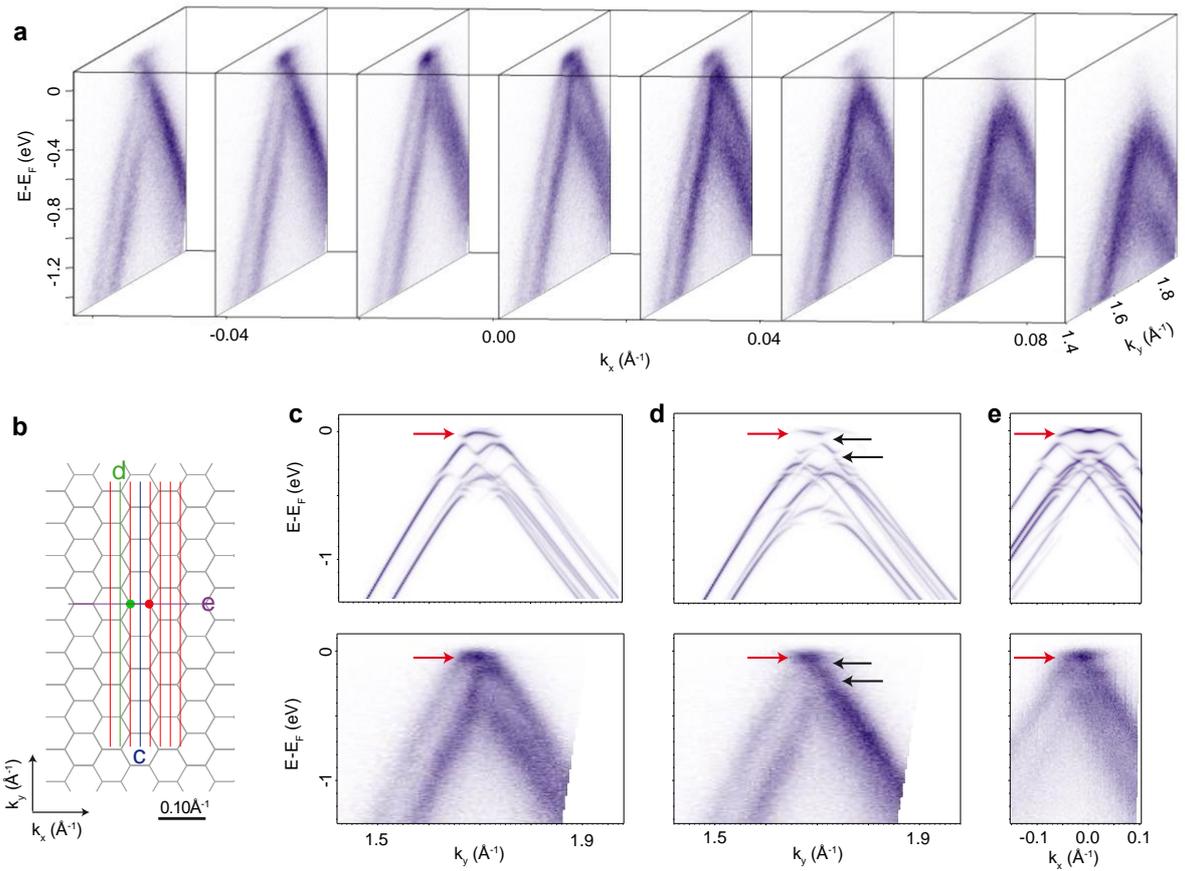

**Figure 4 | Flat band and hybridization gaps. a** Dispersion plots taken on the grid of vertical lines indicated in **b**. Green and red dots in **b** mark the $K_{1,2}$ points of the twisted graphene layers. **c-e** Comparison of individual cuts with calculations of the spectral weight distribution. The latter use a Lorentzian broadening of 10 meV. The flat band and multiple hybridization gaps are marked by red and black arrows, respectively.